\documentclass[review]{elsarticle}
\usepackage{amsmath,amssymb}
\usepackage{graphicx}
\usepackage{hyperref}

\journal{Journal of \LaTeX\ Templates}

\begin{document}

\begin{frontmatter}

\title{Generalized Seniority Schmidt Model and the g-factors in Semi-magic Nuclei}

\author{Bhoomika Maheshwari}
\corref{mycorrespondingauthor}

\address{Department of Physics, Faculty of Science, University of Malaya, Kuala Lumpur, Malaysia}
%
\cortext[mycorrespondingauthor]{Corresponding author}
\ead{bhoomika.physics@gmail.com}
\author{Ashok Kumar Jain}
\address{Amity Institute of Nuclear Science and Technology, Amity University Uttar Pradesh, India}

\begin{abstract}

We have recently applied the generalized seniority approach successfully to explain the B(E1)/B(E2)/B(E3) properties of the semi-magic nuclei. In the present paper, we extend this approach to the Schmidt model as Generalized Seniority Schmidt Model and calculate the g-factors of the various seniority states in the semi-magic nuclei. We find that the magnetic moments and the g-factors do show a particle number independent behavior in multi-j configurations, as expected in the seniority scheme. The calculated results explain the experimental trends quite well. We find that the g-factors of all the seniority states arising from a given multi-j configuration for identical nucleons is equal to the g-factor of the seniority $v=1$ state from that configuration. Also, the g-factors are found to be a sensitive probe for fixing the multi-j configuration, which are fully consistent with the configurations assigned to explain the B(EL) properties in our previous works. We are also able to make definite predictions for many cases.

\end{abstract}

\begin{keyword}
\texttt{Sn-isotopes, Pb-isotopes, $N=82$ isotones, Generalized Seniority, g-factor, Schmidt model, $2_1^+$ states, Seniority isomers}
\end{keyword}

\end{frontmatter}


\section{Introduction}

Recent experimental advances in reaching the neutron drip line provide us new opportunities to test the theoretical estimates, examine new data patterns, and explore the underlying physics ~\cite{jain2015,maheshwari2015,maheshwari2016}. We have recently proposed a simple generalized seniority scheme for multi-j degenerate orbitals, which has been quite successful in obtaining new results and explaining several unresolved features of semi-magic nuclei ~\cite{jain2015, maheshwari2015, maheshwari2016, maheshwarinpa2016, jain2017, jainphysica2017, maheshwari2017}. For example, a new kind of seniority isomers which decay by odd multipole transitions have been found and the related seniority selection rules established for the first time ~\cite{maheshwari2016, maheshwarinpa2016, jain2017}. In the present paper, we extend the generalized seniority calculations to calculate the g-factors of semi-magic nuclei. The magnetic dipole moment (or, the g-factor) of a nuclear state depends on the orbital and spin angular momentum contributions of the contributing protons and neutrons. So the configuration mixing and the single particle structure near the Fermi surface affect the g-factor values significantly. 

Tin isotopes $(Z=50)$ present a fertile ground to investigate the electromagnetic properties, since it is known as the longest semi-magic isotopic chain from the doubly magic $^{100}$Sn to the next doubly magic $^{132}$Sn and beyond. We have recently investigated the first excited $2^+$ states for Sn isotopes which exhibit a particle number independent energy variation throughout the chain and successfully explained the long standing puzzle of two asymmetric B(E2) parabolas with a minimum in the middle of the chain in terms of generalized seniority ~\cite{maheshwarinpa2016}.
 
Several theoretical attempts have been made to understand the g-factor trend of the first excited $2^+$ states in Sn isotopes. For example, the theoretical groups of Terasaki et al.~\cite{terasaki2002}, Brown et al.~\cite{brown2005}, Ansari et al.~\cite{ansari2007} and Jia et al.~\cite{jia2007} have independently studied and predicted the g-factors (magnetic moments) of the neutron-rich Sn isotopes. The quasiparticle random phase approximation (QRPA) calculations of Terasaki et al.~\cite{terasaki2002} are the only calculations, which predicted a positive g-factor for the first excited $2^+$ state in $^{128}$Sn. The shell model calculations by Brown et al.~\cite{brown2005} predicted a constant negative g-factor for the first excited $2^+$ states in $^{124-130}$Sn isotopes. The relativistic QRPA (RQRPA) calculations by Ansari and Ring~\cite{ansari2007} supported a decreasing g-factor trend on going from $^{112}$Sn to $^{130}$Sn, in line with the experimental data. Jia et al.~\cite{jia2007} used a Nucleon Pair Approximation (NPA) to the shell model and predicted near-constant negative g-factors for n-rich even-A Sn isotopes that agree with the shell model calculations of Brown et al.~\cite{brown2005}. Besides this, recent papers by Jiang et al.~\cite{jiang2014}, Allmond et al.~\cite{allmond2015}, Kumbartzki et al. ~\cite{kumbartzki2016} and Robinson et al. ~\cite{robinson2017} have discussed and compared the g-factor trend of the first excited $2^+$ states in Sn isotopes. Different physics reasons have been pointed out for different cases. Still, the issue of the sign of the g-factors in many cases persisted. To conclude, a common physical argument is not able to explain the full trend. More investigations, both theoretical as well as experimental, are hence needed to clear the physics picture. 
 
The experimental information, however, still remains incomplete before $^{112}$Sn and beyond $^{128}$Sn. The g-factors of the first excited $2^+$ states in the stable and even-even Sn isotopes were measured by Hass et al.~\cite{hass1980}; they reported a positive and almost zero value of the g-factor in the case of $^{118}$Sn and negative values for the neighboring nuclei $^{116,120}$Sn on the either side. In general, the trend of the measured magnetic moments and the related g-factors in the stable Sn isotopes runs from positive values for the lighter isotopes to negative values for the heavier isotopes for the first $2^+$ states. The key feature is that the g-factors show a change in behavior near the middle of the shell. This change appears quite consistent to the change in the B(E2) values near the middle of the shell as discussed in our previous work\cite{maheshwarinpa2016}. This change in B(E2) values has been attributed to the different configurations before and after the middle; the $g_{7/2}$ contributes dominantly before the middle while the $h_{11/2}$ takes over the dominating role after the middle. However, the generalized seniority remains constant as $v=2$ throughout the chain for these states ~\cite{maheshwarinpa2016, morales2011}. 

In this paper, we explain the g-factor trends for the first excited $2^+$ states in the Sn isotopes by using the same orbital configurations as used for explaining the B(E2) values in the generalized seniority approach~\cite{maheshwarinpa2016}. The configurations suggested by generalized seniority are merged with the Schmidt model giving us a Generalized Seniority Schmidt Model (GSSM).

Using the GSSM, we obtain and compare the g-factor trends for the high spin ${11/2}^-$ states and the ${10}^+$ isomers in the Sn isotopes. The calculated trends are in line with the experimental data. Predictions for the ${27/2}^-$ isomers have also been made. The paper also presents a comparative study of the g-factor trends in the $Z=50$, and $Z=82$ isotopic chains, with the $N=82$ isotonic chain. We have already discussed similar B(E2) trends of ${10}^+$ and ${27/2}^-$ isomers, understood as generalized seniority $v=2$ and $v=3$ states in the $Z=50$ and $N=82$ chains, and compared them to the ${12}^+$ and ${33/2}^+$ isomers respectively, which are generalized seniority $v=2$ and $v=3$ states in the Pb isotopes~\cite{jain2017}.
 
In short, this paper presents a comparative study of g-factors in different semi-magic chains for various isomers and other excited states. The paper is organized as follows. In Sec. 2, a brief discussion of the generalized seniority scheme used by us, and the formulas for g-factors are given in terms of GSSM. In Sec. 3, we present our g-factor calculations for various states in the $Z=50$, $N=82$, and $Z=82$ chains including the construction of subspace chosen for the multi-j configurations. Sec. 4 summarizes the present work.

\section{The Generalized Seniority Schmidt model and the g-factors}

The seniority scheme, first introduced by Racah~\cite{racah} for single-j shell, is widely used for studying the spectroscopic properties in the semi-magic nuclei~\cite{talmi1993, isacker, isacker1}. In simple terms, seniority is defined as the number of unpaired nucleons in the given j-shell. Kerman~\cite{kerman} and Helmers~\cite{helmers} proposed a simpler description of seniority in terms of a pair creation operator defined as $S_j^+=\frac{1}{2} \sum (-1)^{(j-m)} a_{jm}^+ a_{j,-m}^+ $ and the related quasi-spin algebra. Therefore, a state of seniority $v$ in pure-j configuration, $j^v$, can be defined as $ S_j^- |j^v v J M > =0 $  where $S_j^-$ is the pair annihilation operator, a Hermitian conjugate of the pair creation operator. On applying the Wigner-Eckart theorem to the quasi-spin algebra, it may be inferred that the odd Hermitian tensor operators behave as a quasi-spin scalar while the even Hermitian tensor operators behave as a zero component quasi-spin vector. As a consequence, the reduced matrix elements of electric even tensor multipole operators in $j^n$ configuration ($n$ particles in $j$-shell) can be related to the reduced matrix elements in $j^v$ configuration as follows~\cite{talmi1993}: \\
a. For seniority conserving $\Delta v=0$ transitions
\begin{eqnarray}
\langle {j}^n v J_f ||\sum_i r_i^L Y^{L}|| {j}^n v J_i \rangle = \Bigg[ \frac{\Omega-n}{\Omega-v} \Bigg] \langle {j}^v v J_f ||\sum_i r_i^L Y^{L}|| {j}^v v J_i \rangle
\end{eqnarray}
b. For seniority changing $\Delta v=2$ transitions
\begin{eqnarray}
\langle {j}^n v J_f ||\sum_i r_i^L Y^{L}|| {j}^n v\pm 2 J_i \rangle  = \Bigg[ \sqrt{\frac{(n-v+2)(2\Omega+2-n-v)}{4(\Omega+1-v)}} \Bigg] \nonumber\\ \langle {j}^v v J_f ||\sum_i r_i^L Y^{L}|| {j}^v v\pm 2 J_i \rangle 
\end{eqnarray}
Hence, the reduced electric even tensor transition probabilities show a parabolic trend with a dip or peak in the middle of the shell depending upon the seniority conserving/seniority changing transition. On the other hand, the magnetic multipole transition operator $O(ML)$ behaves as odd tensor in the single-j scheme for the electromagnetic transitions, thus conserving parity. Therefore, the reduced matrix elements for such transitions can be written as: 
\begin{eqnarray}
\langle {j}^n v J_f ||O(ML)|| {j}^n v J_i \rangle = \langle {j}^v v J_f ||O(ML)|| {j}^v v  J_i \rangle
\end{eqnarray}
This can further result in the particle number independent behavior of magnetic moments for identical nucleons~\cite{talmi1993}, since the matrix elements of magnetic dipole moments in $j^n$ configuration can be reduced to the matrix elements of magnetic dipole moments in $j^v$ configuration without $'n'$ dependence as follows: 
\begin{eqnarray}
\langle {j}^n |\hat{\mu} | {j}^n \rangle = \langle {j}^v |\hat{\mu}| {j}^v  \rangle
\end{eqnarray}
Therefore, the magnetic moments for a given seniority $v$ state will be enough to know the magnetic moments of the other isotopes/isotones having the same seniority state in $j^v$ configuration. In this way, the magnetic moment $(\mu)$ of $n$ identical nucleons in a single-j orbital giving rise to total angular momentum $J$, can be written as

\begin{eqnarray}
\vec{\mu}=\sum_i^n g \vec{j_i} =g \sum_i^n \vec{j_i} = g\vec{J}
\end{eqnarray}

The g-factor is simply the ratio of the magnetic moment $\mu$ and $J$. It is nearly same for all the seniority states arising from $j^n$ configuration. Also, the variation of g-factor is going to be particle number independent as per the seniority scheme. Hence, g-factors of different seniority states will be approximately equal to the g-factor of a single seniority state for any given pure $j^n$ configuration.
 
To deal with more realistic situations in nuclei, it became necessary to incorporate the multi-j environment in the seniority scheme. In 1960’s, Arima and Ichimura~\cite{arima}  introduced for the first time the concept of generalized seniority by taking multi-j degenerate orbitals into account. This generalization can also be achieved in a simpler way by generalizing the definition of the pair creation operator~\cite{kerman1} as $S^+=\sum_j S_j^+ $ where the summation takes care of all the active orbitals in any multi-j situation; the complete details may be found in the book of Talmi~\cite{talmi1993}. Talmi then extended this concept to the many non-degenerate orbitals by defining $S^+=\sum_j \alpha_j S_j^+ $, where $\alpha_j$ are the mixing coefficients~\cite{talmi, shlomo}. We have recently extended the usage of generalized seniority scheme by considering the definition of Arvieu and Moszokowski~\cite{arvieu} as $S^+=\sum_j (-1)^{l_j} S_j^+ $ . We have considered the situation of total number of $n$ particles in multi-j orbitals where $n=\sum_j n_j$ and the seniority in single-j changes to the term generalized seniority $v$ in multi-j. We can write a state of generalized seniority $v$ in multi-j $\tilde{j}^v$ configuration as $ S^- |\tilde{j}^v v J M > =0 $ by defining $\tilde{j}=j \otimes j' ....$ with the pair degeneracy of $\Omega=\frac{1}{2} (2\tilde{j}+1)=\sum \frac{1}{2} (2j+1)$. Due to the multi-j environment, we can not stick to the integer nature of $n_j$ and rather explain it in terms of actual neutron occupancies. The occupancies in any j-orbital may/may not be an integer; however, the sum of total occupancies (the total number of particles $n$ and the corresponding generalized seniority $v$ for any state) comes out to be an integer. This also hints towards the quasi-particle picture of the nucleons in multi-j, where we can not ignore the possibility of shared occupancy (or configuration mixing). In this way, the generalized seniority may become a probe to highlight the role of configuration mixing. By using the quasi-spin algebra and tensor properties, we have established the following seniority reduction formulae for electromagnetic transitions~\cite{maheshwari2016} \\
a. For generalized seniority conserving $\Delta v=0$ transitions
\begin{eqnarray}
\langle \tilde{j}^n v l J_f ||\sum_i r_i^L Y^{L}|| \tilde{j}^n v l' J_i \rangle = \Bigg[ \frac{\Omega-n}{\Omega-v} \Bigg] \langle \tilde{j}^v v l J_f ||\sum_i r_i^L Y^{L}|| \tilde{j}^v v l' J_i \rangle
\end{eqnarray}
b. For generalized seniority changing $\Delta v=2$ transitions
\begin{eqnarray}
\langle \tilde{j}^n v l J_f ||\sum_i r_i^L Y^{L}|| \tilde{j}^n v\pm 2 l' J_i \rangle  = \Bigg[ \sqrt{\frac{(n-v+2)(2\Omega+2-n-v)}{4(\Omega+1-v)}} \Bigg] \nonumber\\ \langle \tilde{j}^v v l J_f ||\sum_i r_i^L Y^{L}|| \tilde{j}^v v\pm 2 l' J_i \rangle 
\end{eqnarray}

The above relations look quite similar to those of single-j; however, a major difference arises due the possibility of parity change in the multi-j case, provided the given set of multi-j orbitals has at least one orbital of different parity; $l$ and $l’$ show the parities of final and initial states. In this situation, the reduced electric transition probabilities will show a parabolic behavior irrespective of the nature of involved tensor in the transition (for both even and odd tensors). Such possibilities have further resulted in a new set of generalized seniority selection rules for seniority isomerism. On this basis, we have successfully established a new category of $E1$ decaying seniority isomers in contrast to the general belief of having only $E2$ seniority isomers~\cite{maheshwari2016}. On the other hand, the magnetic transition probabilities can also be having contributions from both even and odd tensor operators depending upon the parities of any given electromagnetic transition. Furthermore, the magnetic moments (corresponding to $M1$ operator) will exhibit a particle number independent behavior in generalized seniority scheme also, similar to the seniority scheme. The corresponding matrix elements of magnetic dipole moments in multi-j $\tilde{j}$ can simply be written as:
\begin{eqnarray}
\langle \tilde{j}^n |\hat{\mu} | \tilde{j}^n \rangle = \langle \tilde{j}^v |\hat{\mu}| \tilde{j}^v  \rangle
\end{eqnarray}
We can, therefore, write the magnetic moment of identical nucleons in the mixed-j configuration $\tilde{j}^n$ as 

\begin{eqnarray}
\vec{\mu}=g \sum_i^n {\vec{\tilde{j}}_i} = g\vec{J}
\end{eqnarray}

Here, $\tilde{j}=j \otimes \j'....$ represents the multi-j configuration having the sum of shared occupancies as $n$, the total particle number. Therefore, the g-factors of a multi-j configuration also exhibit a particle number independent behavior, similar to the single-j case. As a result, the g-factors of all the states arising from a given multi-j configuration having identical nucleons must be equal to the g-factor of a single nucleon (the seniority $v=1$ state) arising from the same multi-j configuration. We can, therefore, test this for various states in semi-magic nuclei having identical nucleons. If this is true, the effective interaction will be nearly diagonal in the generalized seniority scheme. 

As we have already discussed and verified the goodness of generalized seniority for various isomers and other excited states in Sn isotopes, $N=82$ isotones and Pb isotopes ~\cite{maheshwari2016, maheshwarinpa2016, jain2017, jainphysica2017, maheshwari2017}, the present paper reports the results of magnetic moments (or, g-factors) from generalized seniority scheme for the same cases. Talmi has presented and discussed the g-factor trend for single-j seniority scheme consisting of $h_{9/2}$ orbital in $N=126$ isotones by fitting the experimental data of magnetic moments~\cite{talmi1993}. However, no similar results are known for generalized seniority scheme consisting of multi-j orbitals. We, therefore, reach to the following conclusion: If the states are of good generalized seniority then they must have a constant and particle number independent behavior of g-factor throughout any given multi-j configuration. 

In this paper, we calculate the g-factors by merging the idea of generalized seniority with the well-known Schmidt model and term it as Generalized Seniority Schmidt Model. Schmidt model is an extreme single-particle model to calculate the trend of g-factors in odd-A nuclei, particularly near the magic numbers. The g-factor expressions are well known for single-j as    
\begin{eqnarray}
g  =& \frac{1}{j} \Bigg[ {\frac{1}{2} g_s}+ (j- \frac{1}{2}) g_l \Bigg]; j=l+\frac{1}{2}\nonumber\\ 
   =& \frac{1}{j+1} \Bigg[ -\frac{1}{2} g_s + (j+ \frac{3}{2}) g_l \Bigg]; j=l-\frac{1}{2} 
\end{eqnarray}  

where $g_s$ and $g_l$ are taken to be 5.59 n.m. and 1 n.m. for protons, while -3.83 n.m. and 0 n.m. for neutrons, respectively. We extend these expressions in GSSM, by using the definition $\tilde{j} = j \otimes j'....$  for multi-j situations and replace $j$ by $\tilde{j}$, so that the pair degeneracy of mixed configuration is given by $\Omega=\sum{\frac{1}{2} (2\tilde{j}+1)}$ corresponding to the generalized seniority $v$, as follows:
\begin{eqnarray}
g  =& \frac{1}{\tilde{j}} \Bigg[ {\frac{1}{2} g_s}+ (\tilde{j}- \frac{1}{2}) g_l \Bigg]; \tilde{j}=\tilde{l}+\frac{1}{2}\nonumber\\ 
   =& \frac{1}{\tilde{j}+1} \Bigg[ -\frac{1}{2} g_s + (\tilde{j}+ \frac{3}{2}) g_l \Bigg]; \tilde{j}=\tilde{l}-\frac{1}{2} 
\end{eqnarray}
 
As we already know the generalized seniority $v$ and the related configurations for the chosen cases from our previous works ~\cite{maheshwari2016, maheshwarinpa2016, jain2017, jainphysica2017, maheshwari2017}, we use the same configurations to calculate the g-factors of various excited states and isomers in semi-magic nuclei, particularly in $Z=50$, $Z=82$ and $N=82$ chains of nuclei, through GSSM expressions. We call it as GSSM, since the expressions are simply from Schmidt model along with the configuration mixing as suggested by generalized seniority scheme. A comparison with the results of the Schmidt model (pure-j) is also presented.

\section{Results and Discussion}

\begin{figure}
\includegraphics[width=13cm,height=11cm]{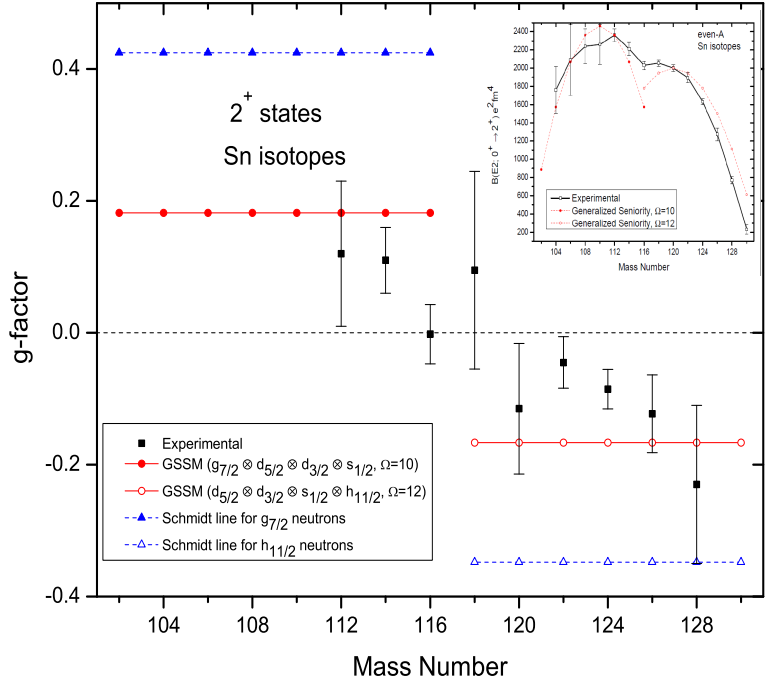} 
\caption{\label{fig:fig1}(Color online) Comparison of the experimental data~\cite{stone2014} and the GSSM calculated g-factor trends for the first excited $2^+$ states in Sn isotopes. Note that the experimental points represent a weighted average of multiple measurements in many cases. The GSSM calculations use different configuration before ($\Omega=10$) and after ($\Omega=12 $) the middle at $^{116}$Sn, as suggested by generalized seniority scheme while explaining the B(E2) trend for these states ~\cite{maheshwarinpa2016} (as shown in the inset). Schmidt lines of $g_{7/2}$ and $h_{11/2}$ neutrons have also been shown before and after the middle, respectively, for comparison.} 
\end{figure}

The g-factors and half-lives of the first excited $2^+$ states in the stable, neutron-deficient Sn isotopes have been measured, but there is little information on the neutron-rich Sn isotopes. We present in Fig.~\ref{fig:fig1}, a comparison of the experimental and the calculated g-
factor trends for the $2^+$ states. The measured values exhibit a trend from positive values for the lighter isotopes (before the middle $<$ $^{116}$Sn) to negative values for the heavier isotopes (after the middle $>$ $^{116}$Sn), and nearly zero value at the middle ($^{116}$Sn). All the experimental data have been taken from Stone's compilation~\cite{stone2014}, and a weighted averaged value obtained in case of multiple measurements.

We have already shown earlier ~\cite{maheshwarinpa2016} that the two parabolas in the B(E2) plot with a dip in the middle are due to a distinct change in the configuration before and after the middle in the generation of the $2^+$ states in the Sn isotopes. The shell model calculations also reproduce the observed trend and support this choice of sub-spaces. Resolution of this long standing puzzle of nuclear structure physics in ~\cite{maheshwarinpa2016} receives further support from the g-factor trends presented in this work. Accordingly, the g-factors are also expected to behave differently before and after the middle of the active valence space. The experimental data does show different yet a nearly particle number independent behavior before and after the mid-shell near $N=66$, $^{116}$Sn.

We, therefore, calculate the g-factor by using the equation (3) for the complete chain of Sn isotopes with two multi-j configurations (in line with the choices made in our previous paper ~\cite{maheshwarinpa2016}) corresponding to $\Omega=10$ and $\Omega=12$ for $\tilde{j}=g_{7/2} \otimes d_{5/2} \otimes s_{1/2} \otimes d_{3/2}=19/2$ and $\tilde{j}= d_{5/2} \otimes s_{1/2} \otimes d_{3/2} \otimes h_{11/2}=23/2$, respectively. The GSSM estimates are plotted in Fig. ~\ref{fig:fig1} along with the experimental data. We emphasize that the GSSM results follow the experimental trends correctly. The multi-j configurations used in explaining the g-factor trends are the same as used in explaining the B(E2) trends for the $2^+$ states in Sn isotopes ~\cite{maheshwarinpa2016} as shown in the inset of Fig.~\ref{fig:fig1}. The B(E2) results obtained from generalized seniority are also supported by the shell model calculations carried out by using the same set of orbitals. It may also be noted that $g_{7/2}$ and $h_{11/2}$ are the highest spin orbitals in this shell model space. The $g_{7/2}$ orbital lies lower in energy and gets filled up first. The $h_{11/2}$ orbital lies higher and begins to fill up later. This is the main reason for $g_{7/2}$ being dominant for lighter Sn isotopes and $h_{11/2}$ being dominant in Sn isotopes after the mid-shell. The Schmidt single particle g-factors of pure $g_{7/2}$ neutrons (before the middle) and $h_{11/2}$ neutrons (after the middle) come out to be 0.425 $n.m.$ (positive) and -0.348 $n.m.$ (negative), respectively, which are quite different from the experimental values, as may be seen in Fig.~\ref{fig:fig1}. 

The multi-j configuration $\tilde{j}$ has been considered to be originating from $(\tilde{l}-1/2)$ before the middle since $g_{7/2}$ dominates. On the other hand, it has been considered to be originating from $(\tilde{l}+1/2)$ after the middle since $h_{11/2}$ dominates. Any other choice takes us further away from the experimental trends. This choice also supports the dominance of the chosen orbitals in multi-j configuration before and after the middle. Similarly, any other choice of subspaces such as $\Omega=7$ and $\Omega=9$ tried in our previous paper ~\cite{maheshwarinpa2016} is not able to explain the experimental B(E2) data. The results, hence, validate the chosen choice of multi-j configurations in the first excited $2^+$ states for Sn isotopes having the generalized seniority as $v=2$, in line with the previous interpretations~\cite{maheshwarinpa2016}. Since g-factors reflect the location of nucleons at the Fermi surface, the positive g-factor values in $^{112,114}$Sn may be understood in terms of the dominance of the $g_{7/2}$ neutrons before the middle. On the other hand, the $h_{11/2}$ neutrons result in negative g-factor values for the heavier isotopes towards $^{132}$Sn. The positive g-factor value (measured) at $^{118}$Sn has quite large experimental error bar and is a fit case for re-measurement. In fact many cases have a large experimental error bar and need a fresh measurement.

Many theoretical attempts have been made to understand and predict the g-factor trend of the first excited $2^+$ states in Sn isotopes in the last decade ~\cite{terasaki2002, brown2005, ansari2007, jia2007, jiang2014, allmond2015, kumbartzki2016, robinson2017}. Our calculated results are in line with the shell model predictions of a constant negative g-factor for $^{124-130}$Sn isotopes by Brown et al.~\cite{brown2005} and in contrast to the QRPA calculations of Terasaki et al.~\cite{terasaki2002}, particularly at $^{128}$Sn, where they predict a positive g-factor. On the other hand, Ansari and Ring~\cite{ansari2007} predicted a decreasing g-factor trend on going from $^{112}$Sn to $^{130}$Sn by RQRPA calculations. Our calculations partially support the same by showing a change in the middle (of the valence space) from the positive g-factor values to the negative g-factor values. However, we expect a nearly particle number independent behavior before and after the middle, which are in line with the nucleon pair approximation calculations of Jia et al.~\cite{jia2007} and the shell model calculations of Brown et al.~\cite{brown2005}. We have explained and also predicted the g-factors in lighter Sn isotopes by using the configuration mixing of $\tilde{j}=g_{7/2} \otimes d_{5/2} \otimes s_{1/2} \otimes d_{3/2}=19/2$. This is in line with the recent large scale shell model interpretations of Kumbartzki et al. ~\cite{kumbartzki2016} where they obtained the best fit to explain g-factors for $^{110,112,114}$Sn without $h_{11/2}$ orbital. The success of this simple model is quite interesting especially when no single theory is able to explain the full trend.    

\begin{figure}
\includegraphics[width=13cm,height=11cm]{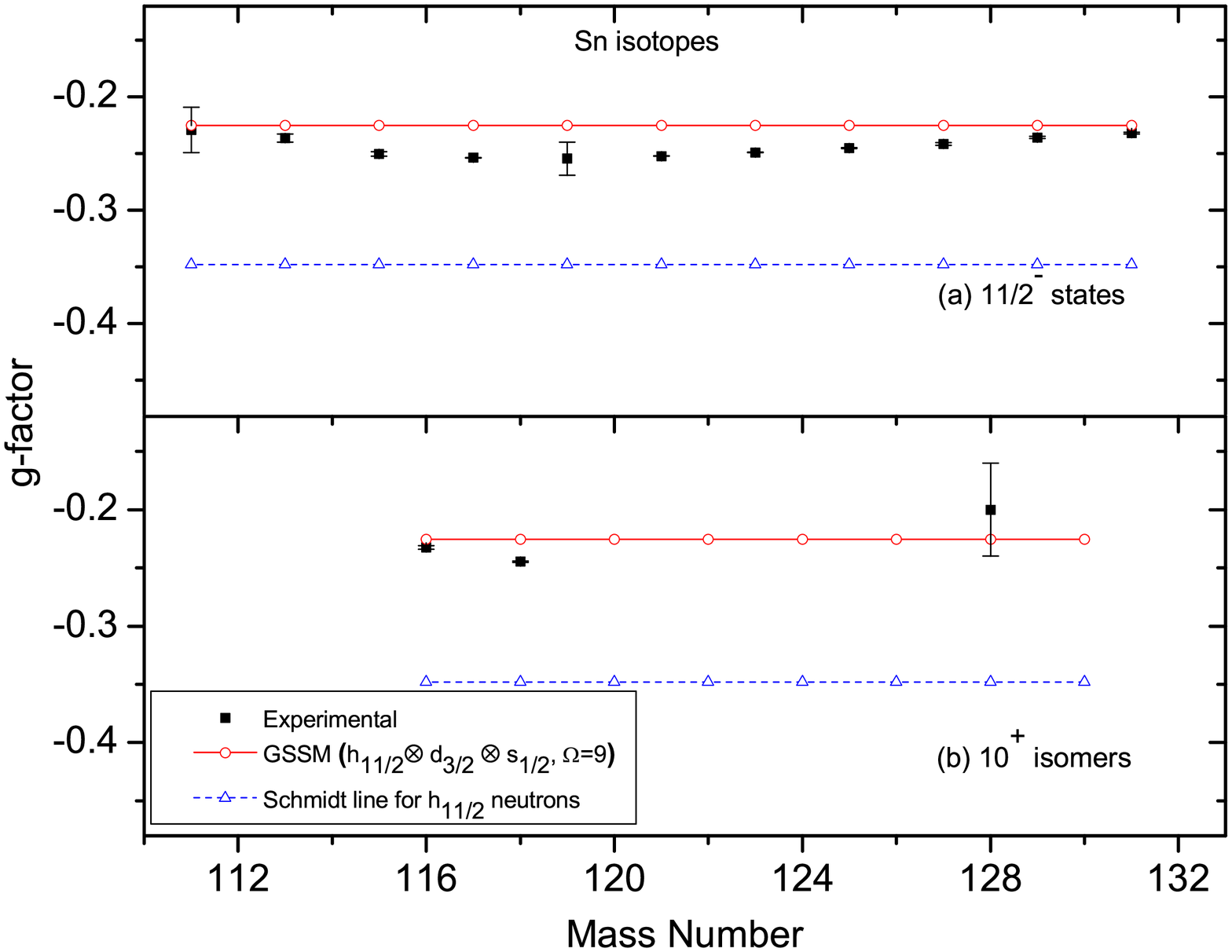}
\caption{\label{fig:fig2}(Color online) The experimental data~\cite{stone2014} and the GSSM calculated g-factor trends for (a) ${11/2}^-$ states and (b) ${10}^+$ isomers in Sn isotopes, respectively. The GSSM calculations have used $\Omega=9$ corresponding to the $\tilde{j}=d_{3/2} \otimes s_{1/2} \otimes h_{11/2}$ configuration, as suggested by generalized seniority scheme in our previous work~\cite{jain2017}. Schmidt lines for $h_{11/2}$ neutrons are shown for comparison.} 
\end{figure}

Next, we present in Fig.~\ref{fig:fig2}(a) and ~\ref{fig:fig2}(b), the g-factor trends for the ${11/2}^-$ states and ${10}^+$ isomers in Sn isotopes, respectively. We can clearly see a particle number independent behavior of g-factors for both the ${11/2}^-$ states and ${10}^+$ isomers, as expected from the seniority and generalized seniority scheme. The ${11/2}^-$ states are expected to be seniority $v=1$ states with $h_{11/2}$  unique-parity orbital configuration of the $50-82$ neutron valence space. But it seems to have a mixed wave function since the Schmidt g-factors for $h_{11/2}$  neutrons come out to be -0.348 $n.m.$, which is quite far from the experimental values as shown in Fig. ~\ref{fig:fig2}(a). We calculate the g-factor trend for the ${11/2}^-$ states by the GSSM along with the same configuration mixing  as used for the ${10}^+$ isomers~\cite{jain2017}. Interestingly, the calculated results come quite close to the experimental data, as shown in Fig. ~\ref{fig:fig2}(a). This strongly hints towards a picture having shared occupancy of $v=1$ from $\tilde{j}=d_{3/2} \otimes s_{1/2} \otimes h_{11/2}$ configuration for the ${11/2}^-$ states in the Sn isotopes. 
\begin{figure}
\includegraphics[width=13cm,height=11cm]{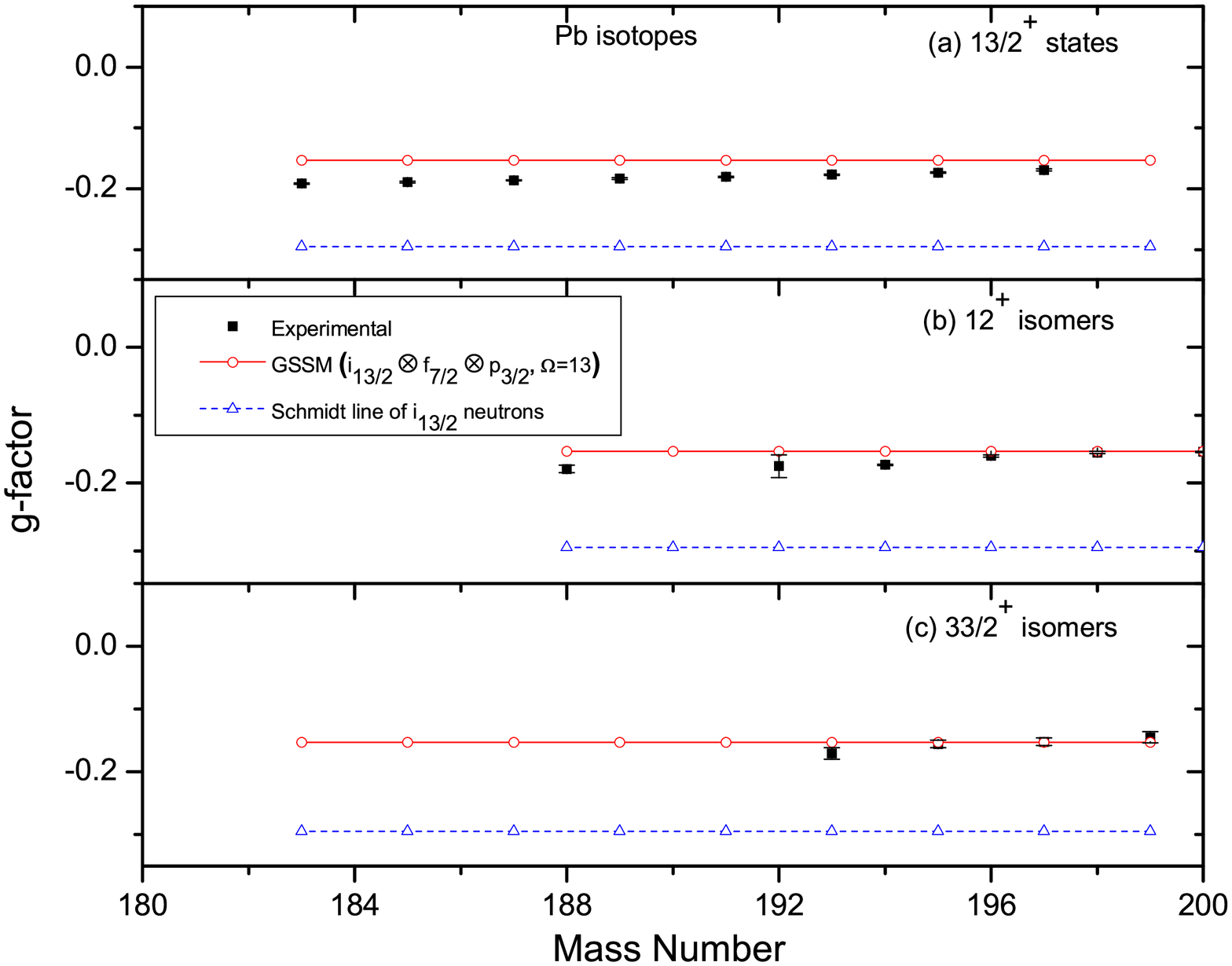} 
\caption{\label{fig:fig3}(Color online) The experimental data ~\cite{stone2014} and the GSSM calculated g-factor trends for the (a) ${13/2}^+$ states, (b) ${12}^+$ isomers and (c) ${33/2}^+$ isomers in Pb isotopes, respectively. The GSSM calculations have used $\Omega=13$ corresponding to the $\tilde{j} = p_{3/2} \otimes f_{7/2} \otimes i_{13/2}$ configuration, as suggested by the generalized seniority scheme in our previous work ~\cite{jain2017}. Schmidt lines for $i_{13/2}$ neutrons are shown for comparison.} 
\end{figure}
\begin{figure}
\includegraphics[width=13cm,height=11cm]{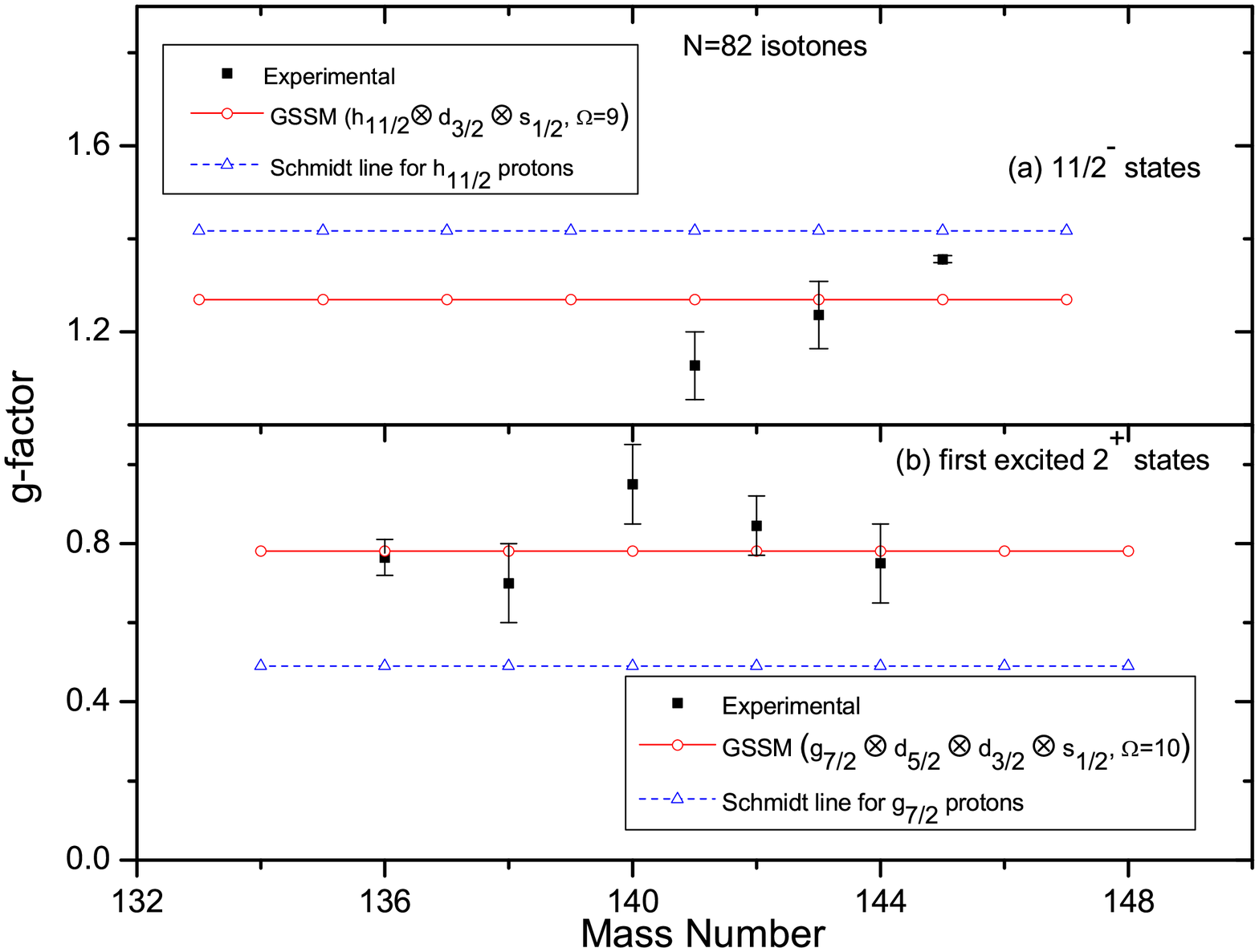}
\caption{\label{fig:fig4}(Color online) The experimental data~\cite{stone2014} and the GSSM calculated g-factor trends for the (a) ${11/2}^-$ states and (b) the first excited $2^+$ states in N=82 isotones, respectively. The GSSM calculations have used $\Omega=9$ and $\Omega=10$ corresponding to $\tilde{j}=d_{3/2} \otimes s_{1/2} \otimes h_{11/2}$ and  $\tilde{j}= g_{7/2} \otimes d_{5/2} \otimes d_{3/2} \otimes s_{1/2}$ configurations, respectively. The Schmidt lines for $g_{7/2}$ protons and $h_{11/2}$ protons are shown, respectively, for the first excited $2^+$ states and the ${11/2}^-$ states.} 
\end{figure}

We present in Fig.~\ref{fig:fig2}(b), GSSM g-factor trend of the ${10}^+$ isomers in Sn isotopes by using the same multi-j configuration, and generalized seniority $v=2$, as used for the B(E2) calculations of these isomers in our previous paper~\cite{jain2017}. We obtain a trend (particle number independent), which matches with the experimental data quite well. Note that the Schmidt g-factor of pure $h_{11/2}$  lies quite lower than the experimental data. It may also be noted that the g-factors of ${10}^+$ states are of the order of the g-factors of ${11/2}^-$ states, which strongly supports the role of configuration mixing in the generation of ${11/2}^-$ states. This is also experimentally evident, as the ${10}^+$ and ${27/2}^-$ isomers closely follow each other in the measured excitation energies, if one puts the ${0}^+$ and ${11/2}^-$ states on equal footing~\cite{jain2017}. To sum up, the ${11/2}^-$ states and the ${10}^+$ isomers can be understood as generalized seniority $v=1$ states and generalized seniority $v=2$ states arising from the same multi-j configuration $\tilde{j}=d_{3/2} \otimes s_{1/2} \otimes h_{11/2}$.  That is why the existence of ${10}^+$ isomers are known from $^{116}$Sn to $^{130}$Sn involving a variation of 14 particles (with few experimental gaps in the g-factor measurements), while pure $h_{11/2}$ leads us to a variation of 12 particles only. The present conclusions are fully in line with the previous interpretations on the reduced transition probabilities~\cite{maheshwari2016,jain2017}. These results further suggest that the realistic effective interaction for describing these states should be nearly diagonal in the generalized seniority scheme. This may also be correlated with the more sophisticated Shell model calculations, if one is able to develop a new realistic effective interaction for the $N=65-82$ valence space consisting of the three $h_{11/2}$ , $d_{3/2}$ and $s_{1/2}$ orbitals with $N=64$ $(Z=50)$ as a core. This also leads us to some predictions. Only three measured values are known for the ${10}^+$ isomers; for the rest of the cases, we expect the g-factor values to be of the same order due to the particle number independent behavior.  	

Measurements of the g-factor values for the ${27/2}^-$ isomers are not available. We, therefore, predict the g-factor values of the ${27/2}^-$ isomers to be of the same order as those of the ${10}^+$ isomers, since both follow similar configuration mixing~\cite{jain2017}. A particle number independent g-factor behavior is also expected for the ${27/2}^-$ isomers, which originate from the generalized seniority $v=3$, $\tilde{j}=d_{3/2} \otimes s_{1/2} \otimes h_{11/2}$ configuration. New experimental measurements are needed to confirm these predictions.

We further present g-factor trends for the ${13/2}^+$ states, the ${12}^+$ isomers and the ${33/2}^+$ isomers of $Z=82$ isotopes in Fig.~\ref{fig:fig3}(a),~\ref{fig:fig3}(b) and ~\ref{fig:fig3}(c), respectively. The particle number independent behavior is clearly visible in all the three cases. We have already understood the ${12}^+$ and ${33/2}^+$ isomers in Pb isotopes as the generalized seniority $v=2$ and $v=3$ states arising from $\tilde{j} = p_{3/2} \otimes f_{7/2} \otimes i_{13/2}$ configuration, where their similar B(E2) parabolic trends have been attributed to the goodness of generalized seniority ~\cite{jain2017}. Since the order of the g-factor values for the three ${13/2}^+$, ${12}^+$ and ${33/2}^+$ states are nearly same, we expect that the ${13/2}^+$ states also have a similar configuration mixing and generalized seniority $v=1$. Similar to the ${11/2}^-$ states in the Sn isotopes, the ${13/2}^+$ states in Pb isotopes correspond to the unique-parity $i_{13/2}$ orbital of $N=82-126$ valence space. Still, it follows the GSSM trend for the g-factor values, See Fig. 3(a).  Note that the Schmidt values for pure $i_{13/2}$ orbital are quite far from the experimental trend.

Similar is the case for the ${12}^+$ and ${33/2}^+$ isomers in even-even and even-odd Pb isotopes, respectively. Note that the multi-j configurations used for the g-factors are the same as those used to describe the B(E2) properties in our earlier paper~\cite{jain2017}. This means that the configuration mixing suggested by generalized seniority is consistent in explaining all the electromagnetic properties for both the Sn isotopes and the Pb isotopes. Generalized seniority thus becomes a unique tool to make predictions for the gaps in the measurement.

It may be noted that the neutrons in $Z=50$ isotopes and the protons in $N=82$ isotones occupy $g_{7/2}$, $d_{5/2}$, $h_{11/2}$, $d_{3/2}$, and $s_{1/2}$ orbitals in the $50-82$ nucleon space. An interesting comparison may therefore be made between the two. The ${10}^+$ and ${27/2}^-$ isomers have already been identified as generalized seniority $v=2$ and $v=3$ states arising from $\tilde{j}=d_{3/2} \otimes s_{1/2} \otimes h_{11/2}$ in the $N=82$ isotones, analogous to the situation in the Sn isotopes. However, we do not have the g-factor measurements to compare in both the cases. We predict the g-factor values for both to be of the order of 1.27 $n.m.$, as calculated from GSSM. New measurements should be carried out to confirm the predictions.

We also exhibit in Fig. ~\ref{fig:fig4}(a) and ~\ref{fig:fig4} (b), the experimental and calculated g-factor trends of the first excited $2^+$ states and ${11/2}^-$ states, respectively, in the $N=82$ isotonic chain. We again find nearly particle number independent behavior, as expected. The data are rather limited for these cases. The multi-j configurations used for the first excited $2^+$ states and the ${11/2}^-$ states are, $\tilde{j}=g_{7/2} \otimes d_{5/2} \otimes s_{1/2} \otimes d_{3/2}=19/2$ and  $\tilde{j}=d_{3/2} \otimes s_{1/2} \otimes h_{11/2}=17/2$, respectively, similar to the Sn isotopes. The GSSM calculated values come close to the experimental trend, while pure Schmidt values lie quite far, as shown in Fig. ~\ref{fig:fig4}(a) and ~\ref{fig:fig4}(b). Note that we compare the Schmidt $g_{7/2}$ values in the case of first excited $2^+$ states, while Schmidt $h_{11/2}$ values in the case of ${11/2}^-$ states in the $N=82$ isotones. The g-factor values are positive and much larger for protons ($N=82$ chain) in comparison to the neutrons ($Z=50$ chain). This is expected as the magnetic moment due to the orbital angular momentum survives for protons resulting in larger values, and hence the positive g-factors. 

It is noteworthy that the g-factors for the ${10}^+$ and ${27/2}^-$ isomers will be of the same order as those of the ${11/2}^-$ states in $N=82$ isotones, similar to the results in Sn isotopes. Since generalized seniority remains a reasonably good quantum number for these states in both the Sn isotopes and the $N=82$ isotones arising from similar multi-j configurations, systematic studies on Sn isotopes helped us in making predictions for $N=82$ isotones. In general, a particle number independent behavior of g-factors is observed for all the excited states and isomers discussed so far, suggesting a near goodness of generalized seniority quantum number for them. This is consistent with our previous interpretations, while studying reduced transition probabilities in various semi-magic nuclei ~\cite{maheshwari2016, maheshwarinpa2016, jain2017, jainphysica2017, maheshwari2017}. The present study hence validates the usage of generalized seniority scheme in semi-magic nuclei. 

\section{Conclusion}

In this paper, we have presented a Generalized Seniority Schmidt Model (GSSM) to calculate the g-factor values of various states and isomers in the semi-magic $Z=50$ Sn isotopes, $N=82$ isotones, and $Z=82$ Pb isotopes. We start by exploring the well-known cases of the first excited $2^+$ states in Sn isotopes. Our calculated results for the g-factors explain the experimental trend quite well, and support the particle number independent behaviour along with a transition in the middle from positive g-factor to the negative g-factor. The transition in the middle has been understood due to the dominance of different multi-j configurations before and after the middle; however, generalized seniority remains constant as $v=2$ throughout the full chain. The positive g-factor before the middle is due to the dominance of $g_{7/2}$ neutrons, while the negative g-factor after the middle is due to the dominance of $h_{11/2}$ neutrons. This explanation is fully consistent with our previous interpretations of the B(E2) values for the first excited $2^+$ states, which explains the anomalous double hump in the B(E2) values - a long standing puzzle [4].

We have also studied the g-factor trends for the ${11/2}^-$ states and the ${10}^+$ isomers in Sn isotopes, where the GSSM calculated results explain the experimental data quite well. We have then presented the g-factor trends for the ${13/2}^+$, ${12}^+$ and ${33/2}^+$ states in the Pb isotopes. Similar trends were expected and found in Sn and Pb isotopes due to the goodness of generalized seniority; however, the involved orbitals are different in both Sn isotopes and Pb isotopes. We have also compared the first excited $2^+$ states and ${11/2}^-$ states in $N=82$ isotones. The neutrons in $Z=50$ and protons in $N=82$ occupy similar set of orbitals and result in similar structural phenomena. Predictions have also been made for the gaps in the data in all the three $Z=50$, $N=82$ and $Z=82$ chains. New experiments are needed to confirm these predictions.

The results are interesting as they show that the g-factors and the related matrix elements are very sensitive to the details of the wave function and configuration mixing. Near goodness of generalized seniority is visible due to the expected particle number independent behavior of g-factor values. For the first time, the properties of the ${11/2}^-$ states in $Z=50$ and $N=82$ chains and the ${13/2}^+$ states in $Z=82$ chain have been attributed to the multi-j configuration. It seems that an effective interaction would be nearly diagonal in generalized seniority for these states. This also highlights the need of designing new realistic effective interactions for Shell model calculations.

\section*{Acknowledgments}

AKJ thanks the Amity Institute of Nuclear Science and Technology, Amity University, NOIDA for financial support to complete this work. BM thanks the University of Malaya for the award of a PDF to continue this work.


\begin{thebibliography}{50}

\bibitem{jain2015}A. K. Jain, B. Maheshwari, S. Garg, M. Patial and B. Singh, Nuclear Data Sheets 128 (2015) 1.
\bibitem{maheshwari2015}B. Maheshwari, A. K. Jain and P. C. Srivastava, Phys. Rev. C 91 (2015) 024321.
\bibitem{maheshwari2016}B. Maheshwari and A. K. Jain, Phys. Lett. B 753 (2016) 122.
\bibitem{maheshwarinpa2016}B. Maheshwari, A. K. Jain and B. Singh, Nucl. Phys. A 952 (2016) 62.
\bibitem{jain2017}A. K. Jain and B. Maheshwari, Nuclear Phys. Review 34 (2017) 73. 
\bibitem{jainphysica2017}A. K. Jain and B. Maheshwari, Physica Scripta 92 (2017) 074004.
\bibitem{maheshwari2017}B. Maheshwari, S. Garg and A. K. Jain, Pramana-Journal of Physics (Rapid Communication) 89 (2017) 75.
\bibitem{terasaki2002}J. Terasaki, J. Engel, W. Nazarewicz, and M. Stoitsov, Phys. Rev. C 66 (2002) 054313.
\bibitem{brown2005}B. A. Brown, N. J. Stone, J. R. Stone, I. S. Towner, and M. Hjorth-Jensen, Phys. Rev. C 71 (2005) 044317.
\bibitem{ansari2007}A. Ansari, and P. Ring, Phys. Lett. B 649 (2007) 128.
\bibitem{jia2007}L. Y. Jia, H. Zhang, and Y. M. Zhao, Phys. Rev. C 75 (2007) 034307.
\bibitem{jiang2014}H. Jiang et al., Phys. Rev. C 89 (2014) 014320.
\bibitem{allmond2015}J. M. Allmond et al., Phys. Rev. C 92 (2015) 041303(R).
\bibitem{kumbartzki2016}G. J. Kumbartzki et al., Phys. Rev. C 93 (2016) 044316.
\bibitem{robinson2017}S. Robinson et al., International Journal of Modern Physics E 26 (2017) 1750053.
\bibitem{hass1980}M. Hass, C. Broude, Y. Niv, A. Zemel, Phys. Rev. C 22 (1980) 97.
\bibitem{morales2011}I. O. Morales, P. Van Isacker, I. Talmi, Phys. Lett. B 703 (2011) 606.
\bibitem{talmi1993}I. Talmi, Simple Models of Complex Nuclei, Harwood Academic (1993).
\bibitem{stone2014}N. J. Stone, www-nds.iaea.org/publications,  INDC(NDS)-0658, Feb. (2014). 
\bibitem{racah}G. Racah, Phys. Rev 63 (1943) 367; Research Council of Israel, Jerusalem, L. Farkas Memorial Volume (1952) 294.
\bibitem{isacker}P. Van Isacker, Phys. Rev. Lett. 100 (2008) 052501.
\bibitem{isacker1}P. Van Isacker, J. Phys.: Conf. Ser. 322 (2011) 012003.
\bibitem{kerman}A. K. Kerman, Ann. Phys. (NY) 12 (1961) 300.
\bibitem{helmers}K. Helmers, Nucl. Phys. 23 (1961) 594.
\bibitem{arima}A. Arima and M. Ichimura, Prog. of Theo. Phys. 36 (1966) 296.
\bibitem{kerman1}A. K. Kerman, R. D. Lawson, and M. H. Macfarlane, Phys. Rev. 124 (1961) 162.
\bibitem{talmi}I. Talmi, Nucl. Phys. A 172 (1971) 1.
\bibitem{shlomo}S. Shlomo, and I. Talmi, Nucl. Phys. A 198 (1972) 82.
\bibitem{arvieu}R. Arvieu, and S. A. Moszokowski, Phys. Rev. 145 (1966) 830.


\end{thebibliography}
\end{document}